\documentclass[letterpaper,american,reprint, aps, prl, superscriptaddress]{revtex4-1}
\usepackage[T2A,T1]{fontenc}
\usepackage[latin9]{inputenc}
\usepackage{textcomp}
\usepackage{amsmath}
\usepackage{amssymb, bm}
\usepackage{graphicx}

\makeatletter

\newcommand{\bra}{\left< }
\newcommand{\ket}{\right>}

\makeatother

\begin{document}

\title{Pressure dependence of ferroelectric quantum critical fluctuations}

\author{M.J. Coak}

\affiliation{Cavendish Laboratory, Cambridge University, J.J. Thomson Ave, Cambridge
CB3 0HE, UK}

\affiliation{Center for Correlated Electron Systems, Institute for Basic Science,
Seoul 08826, Republic of Korea}

\affiliation{Department of Physics and Astronomy, Seoul National University, Seoul
08826, Republic of Korea}

\affiliation{Department of Physics, University of Warwick, Coventry CV4 7AL, UK}

\author{C.R.S. Haines}

\affiliation{Cavendish Laboratory, Cambridge University, J.J. Thomson Ave, Cambridge
CB3 0HE, UK}

\author{C. Liu}

\affiliation{Cavendish Laboratory, Cambridge University, J.J. Thomson Ave, Cambridge
CB3 0HE, UK}

\author{G. G. Guzm\'{a}n-Verri}

\affiliation{Cavendish Laboratory, Cambridge University, J.J. Thomson Ave, Cambridge
CB3 0HE, UK}

\affiliation{Centro de Investigaci\'{o}n en Ciencia e Ingenier\'{i}a de Materiales~(CICIMA), Universidad de Costa Rica, San Jos\'{e}, Costa Rica 11501}

\affiliation{Escuela de F\'{i}sica, Universidad de Costa Rica, San Jos\'{e}, Costa Rica 11501}

\author{S.S. Saxena}

\affiliation{Cavendish Laboratory, Cambridge University, J.J. Thomson Ave, Cambridge
CB3 0HE, UK}

\affiliation{National University of Science and Technology \textquotedblleft MISiS\textquotedblright ,
Leninsky Prospekt 4, Moscow 119049, Russia}

\date{\today}
\begin{abstract}

Phase transitions to a long-range-ordered state driven by a softened phonon mode are ubiquitious across condensed matter physics, but the evolution of such a mode as the system is tuned to or from the transition has never been explictly measured until now. We report for the first time the effect of pressure on the soft mode associated with ferroelectricity in the archetypal quantum critical paraelectric SrTiO$_3$. This is an ideal, clean, model system for exploring these effects, with pressure directly addressing the phonon modes only. We measure and report the effect of quantum critical fluctuations on the pressure and temperature dependence of the ferroelectric soft phonon mode as the system is tuned away from criticality. We show that the mean field approximation is confirmed experimentally. Furthermore, using a self-consistent model of the quantum critical excitations including coupling to the volume strain and without adjustable parameters, we determine logarithmic corrections that would be observable only very close to the quantum critical point. Thus, the mean-field character of the pressure dependence is much more robust to the fluctuations than is the temperature dependence. We predict stronger corrections for lower dimensionalities. The same calculation confirms that the Lydanne-Sachs-Teller relation is valid over the whole pressure and temperature range considered. Therefore, the measured dielectric constant can be used to extract the frequency of the soft mode down to $1.5\,$K and up to $20\,$kbar of applied pressure. The soft mode is observed to stiffen further, raising the low-temperature energy gap and returning towards the expected shallow temperature dependence of an optical mode. This behavior is consistent with the existence of a ferroelectric quantum critical point on the pressure-temperature phase diagram of SrTiO$_3$, which applied pressure tunes the system away from. This work represents the first experimental measurement of the stiffening of a soft phonon mode as a system is tuned away from criticality, a potentially universal phenomenon across a variety of phase transitions and systems in condensed matter physics. 
\end{abstract}
\maketitle

\section{Introduction}

The incipient ferroelectric SrTiO$_3$ has been studied in great detail and is used as a dielectric or an insulating substrate in many applications \citep{Barrett1952,Cowley1964,Samara1966,Mueller1979,Muller1991,Fujishita2016,Atkinson2017} as it possess an anomalously high dielectric constant at low temperatures. SrTiO$_3$ is the clearest example of a quantum paraelectric known. The nature of the quantum paraelectric state has been an open question since Barrett \citep{Barrett1952} first attempted to describe the deviation from Curie-Weiss behavior observed in SrTiO$_3$. Recently, progress has been made \citep{Coak2018} in describing this state but the effect of fluctuations on the pressure dependence has not been considered until this work.

\begin{figure}
\centering{}\includegraphics[width=1\columnwidth]{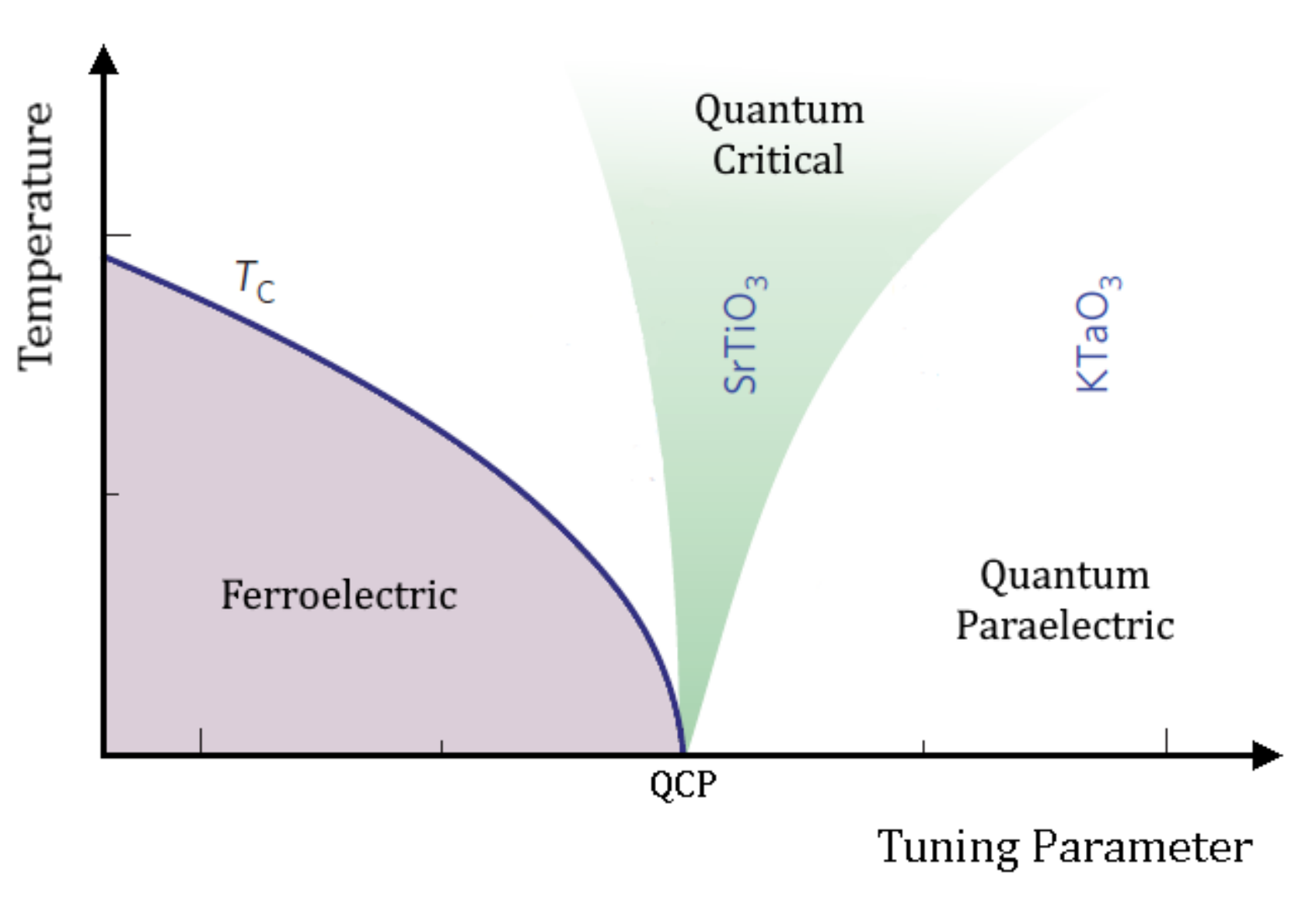}\caption{\label{fig:STO_PhaseDiagram}Predicted phase diagram for SrTiO$_3$ and related compounds \citep{Rowley2014}. The `tuning parameter' can be realized through either chemical doping or substitution, or by applying pressure to tune the frequencies of the phonon oscillators. SrTiO$_3$ sits naturally slightly to the right of a quantum critical point on this diagram, and pressure serves to push it further away from the quantum critical region and towards the behavior of KTaO$_3$.}
\end{figure}

SrTiO$_3$ has a cubic perovskite structure at room temperature with a well-documented antiferrodistortive transition at around 110~K \citep{Shirane1969}, below which the unit cell is tetragonal. The dielectric constant, $\epsilon_{r}$, of SrTiO$_3$ exhibits a classical Curie-Weiss temperature dependence at high temperatures \citep{Mueller1979}, but departs from the Curie-Weiss behavior as the polarization is modified by quantum fluctuations below 50~K, and no ferroelectric ordering is observed. The system remains close to long-range order through a displacive ferroelectric transition, but the transition does not occur in pristine SrTiO$_3$. The polarisation is due to the ionic motion of a transverse optical phonon mode, which softens to approach zero energy at low temperature at the zone center $\Gamma$ point. This mode, commonly described as the `soft mode' or `ferroelectric mode' corresponds to opposing motion of the oxygen octahedra and titanium ions along the c-axis \citep{Cowley1964,Fleury1968}. Inelastic neutron scattering experiments by Yamada et al. \citep{Yamada1969} have shown however that the soft mode is `frozen out' at low temperatures - instead of falling to zero the phonon energy flattens out to a fixed value or gap $\Delta$ as temperature is decreased below 50~K. This means that spontaneous long-range ferroelectricity does not form, but a low-energy excitation into short range charge ordering is available to the system, leading to the very large polarizability and departures from classical predictions seen at low temperatures \citep{Muller1991}. These phenomena have been linked to a postulated ferroelectric quantum critical point (QCP) at the equivalent of a small `negative pressure' on SrTiO$_3$'s pressure-temperature phase diagram (Fig. \ref{fig:STO_PhaseDiagram}) - quantum critical fluctuations of the polarization in the vicinity of this point lead to a modified dependence of the dielectric properties. The application of strain \citep{Uwe1976}, or the substitution of a heavier oxygen isotope to form SrTi$^{18}$O$_3$ \citep{Itoh2000a,Wang2000}, which both have the opposite effect on the system to applying pressure, lead to a ferroelectric phase, consistent with this QCP. Our recent work \citep{Coak2018} has shown through detailed measurements of the dielectric constant under pressure that SrTiO$_3$ is driven away from its quantum critical regime by the application of pressure. The dielectric response was in addition seen to match that seen in KTaO$_3$ at elevated pressures, in agreement with the position on the pressure/doping phase diagram postulated for this material (Fig. \ref{fig:STO_PhaseDiagram}).  Additionally, the dielectric loss in SrTiO$_3$ is dominated by a series of peaks at characteristic temperatures. The largest of these peaks, at approximately 10 K,  is associated with quantum fluctuations. Recent work \citep{Coak2018b} suggests that this peak reflects a localized excitation formed of quantum fluctuations of domain walls.

Critical exponents in the temperature dependence of susceptibility, resistivity and heat capacity have been used extensively to identify and characterize the various states discovered in the heavy fermion systems and incipient transition metal itinerant magnets~\cite{Lonzarich_MagneticElectron}. There are good reasons for this. In particular the power of the temperature critical exponents lies in the incredible precision that temperature can be controlled and measured. This is not in general true for pressure. There are also differences in how pressure and temperature affect the energy scales - for the temperature case, the quantum critical cone is extremely narrow at low $T$ and then widens out, with subtle exponent crossovers, and with quantum critical effects persisting to unintuitively high temperatures. In contrast, a small change in pressure at low temperature - where critical effects are strongest just above the QCP - can push the system out of the quantum critical cone. It is worth understanding why the pressure dependence is mean-field-like - widely assumed but by no means obvious from first principles. For paraelectrics such as STO, the leading effect of the electrostrictive coupling on the zone-centre fluctuations is only to renormalize its anharmonic coefficients \citep{Cowley1964}, as shown later in the Theory section.  Such couplings could also affect the dispersion away from $q=0$ \citep{LinesAndGlass}, but these are higher order effects involving fluctuations of the steric degrees of freedom which we have ignored in our model as their effect is stronger on the acoustic modes \citep{Axe1970}.

In this study the pressure dependence is measured explicitly. The relative paucity of data points compared to the temperature sweeps prohibits detailed numerical analysis. However, by considering a standard microscopic model of the ferroelectric fluctuations coupled to the volume strain and solved within the self-consistent phonon approximation (SCPA) \citep{Arce-Gamboa2017a,Arce-Gamboa2018a}, we reveal that the pressure dependence is indeed mean-field-like, except very close to the QCP where weak logarithmic corrections are expected. As a consequence of this, whilst the crossovers from quantum critical to classical and quantum paraelectric are clear in the temperature exponents, this crossover does not appear in the pressure dependence. The SCPA is equivalent to the self-consistent renormalization method used for itinerant magnets \citep{Loehneysen2007}.

The accompanying changes to the soft mode frequency have not been
studied with pressure and temperature, and are reported here for the
first time, extracted from the temperature and pressure dependence
of the dielectric constant. The Lydanne-Sachs-Teller (LST) relation \citep{Lyddane1941} allows
the dielectric response of a polarizable crystal to be determined
from its phonon spectrum or vice versa. The full relation including
all phonon branches can be written as
\begin{equation}
\frac{\epsilon_{r}}{\epsilon_{\infty}}=\frac{\prod_{i}\omega_{i\,LO}^{2}}{\prod_{i}\omega_{i\,TO}^{2}},\label{eq:LST_Full}
\end{equation}
where $\epsilon_{r}$ denotes the relative permittivity of the material,
$\epsilon_{\infty}=n$ the refractive index or dielectric response
at frequencies well above the relevant phonon frequencies and $\omega_{i\,LO}^{2}$
and $\omega_{i\,TO}^{2}$ are the frequencies of the longitudinal
optical (LO) and transverse optical (TO) of phonon mode $i$ at the
zone center ($q=0$). In SrTiO$_3$ a single TO mode corresponding to
opposite motions of the titanium ion and surrounding oxygen octahedron
is softened at the zone center \citep{Yamada1969} and dominates the
dielectric properties of the system. Following Barker \citep{Barker1966}
and Yamada \citep{Yamada1969}, we simplify the LST relation by assuming
the pressure and temperature dependence of all modes other than the
soft TO and its accompanying LO mode to be insignificant. Over the
20~kbar (2~GPa) range reported in this study, SrTiO$_3$ does not undergo
any structural transitions, changes in lattice parameters are well
under a percent \citep{Guennou2010} and the variation of room-temperature
phonon frequencies has been shown to be linear with pressure up to
a much larger range \citep{Ishidate1992}. The effect of these comparatively
low pressures (several orders of magnitude below the bulk modulus
\citep{Beattie1971}) on SrTiO$_3$ is therefore to cause a small leading-order
perturbation to the lattice and phonons, with the notable exception
of the soft mode. All other optical modes can be expected to tune
with applied pressure in the same manner, and their terms in the LST
relation cancel, leading to
\begin{equation}
\frac{\epsilon_{r}}{\epsilon_{\infty}}=\frac{\omega_{LO}^{2}}{\omega_{TO}^{2}},\label{eq:LST_Reduced}
\end{equation}
where $\omega_{TO}$ and $\omega_{LO}$ denote the ferroelectric soft
mode and its longitudinal equivalent, which is denoted as $E_{g}$
and of energy 148~cm$^{-1}$ by Ishidate et al. \citep{Ishidate1992}.

\section{Experiments}

\subsection{Methods}

High precision capacitance measurements were carried out on single
crystal samples of SrTiO$_3$ from Crystal GmbH with gold electrodes
vacuum evaporated onto the surfaces in a parallel-plate capacitor
geometry. Measurements under extremely hydrostatic pressure conditions
were made possible by the development, in collaboration with CamCool
Research Ltd, of a piston-cylinder clamp cell with miniature shielded
coaxial cables running into the sample region and electrically isolated
from the cell body. This eliminates stray capacitances from the wiring
and allows pF capacitance signals to be measured with stabilities
of one part in a million. The shield conductors of the coaxial cables
were joined together at the sample position and at the measurement
instrument in the standard 2-point capacitance setup. The pressures
in the measurements reported were 0/ambient, 2.4, 3.3, 4.6, 5.2, 6.9,
8.6, 9.6, 11.9, 14.1, 15.7, 17.1, 18.6 and 20.0 kbar, with a typical
uncertainty of $\pm$0.2~kbar. An Andeen-Hagerling 2550A 1~kHz capacitance
bridge was used, at a voltage of 0.1~V. Sample thickness, corresponding
to capacitor plate separation, was 0.5~mm. Measurements were taken
on a modified 1K Dipper cryostat from ICE Oxford, allowing continuous
stable temperature control down to 1.2~K, and an adiabatic demagnetization
refrigerator developed in house capable of reaching 200~mK. Typical
heating or cooling rates were held at 0.01~K per minute to allow
the large thermal mass of the pressure cell to thermally equilibrate.

\subsection{Results}

\begin{figure}
\centering{}\includegraphics[width=1\columnwidth]{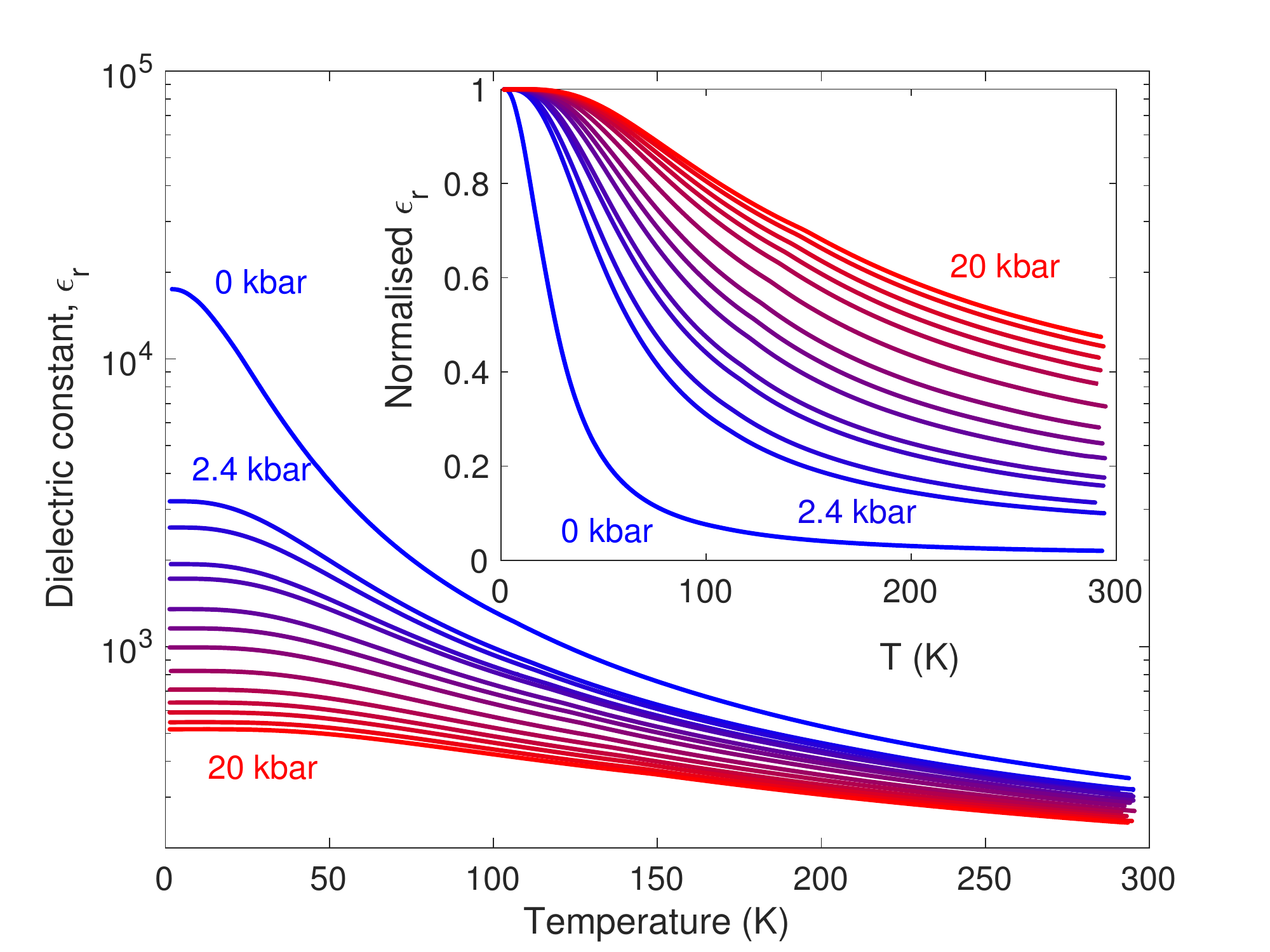}\caption{\label{fig:STO_Data}Temperature dependence of the dielectric constant of SrTiO$_3$ at hydrostatic pressure values from ambient (blue) to 20 kbar (red). The inset shows the same data normalised to their 2 K values, revealing a clear evolution of the shape of the curves as pressure is increased.}
\end{figure}

Measurements were carried out of the dielectric constant of SrTiO$_3$
from room temperature down to 1.5~K at fixed hydrostatic pressures
up to 20.0~kbar. The effect of pressure, as discussed in more detail
in our recent paper \citep{Coak2018}, is to drastically reduce the
magnitude of the low-temperature dielectric constant from \textasciitilde{}18,000
at ambient pressure to \textasciitilde{}500 at 20.0~kbar as the system
is tuned away from a ferroelectric transition and associated quantum
critical behavior - see Fig  \ref{fig:STO_Data}.

\begin{figure*}
\begin{raggedright}
a)\hspace{1.0\columnwidth}b)
\par\end{raggedright}
\centering{}\includegraphics[width=0.4\paperwidth]{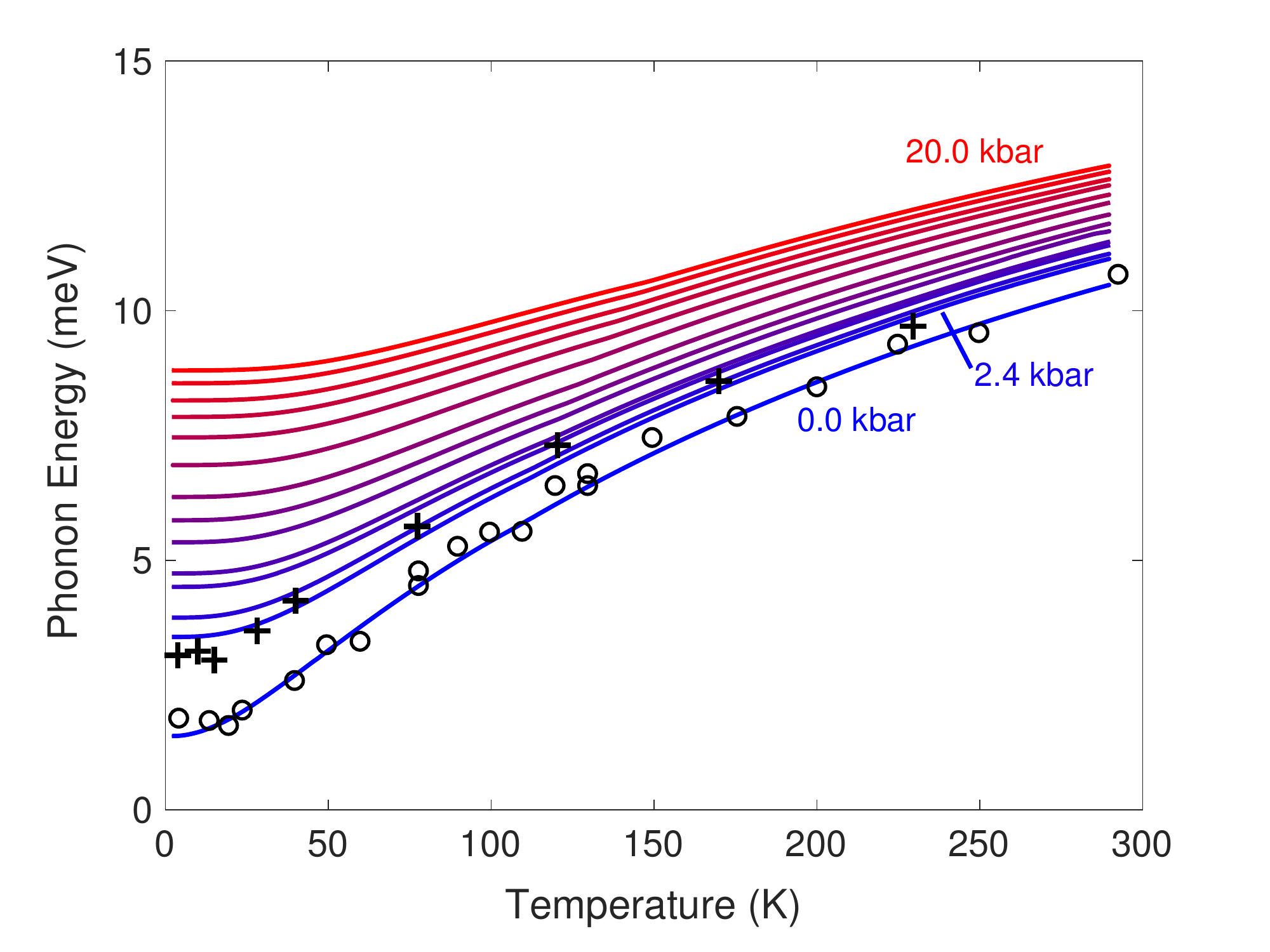}\hfill{}\includegraphics[width=0.4\paperwidth]{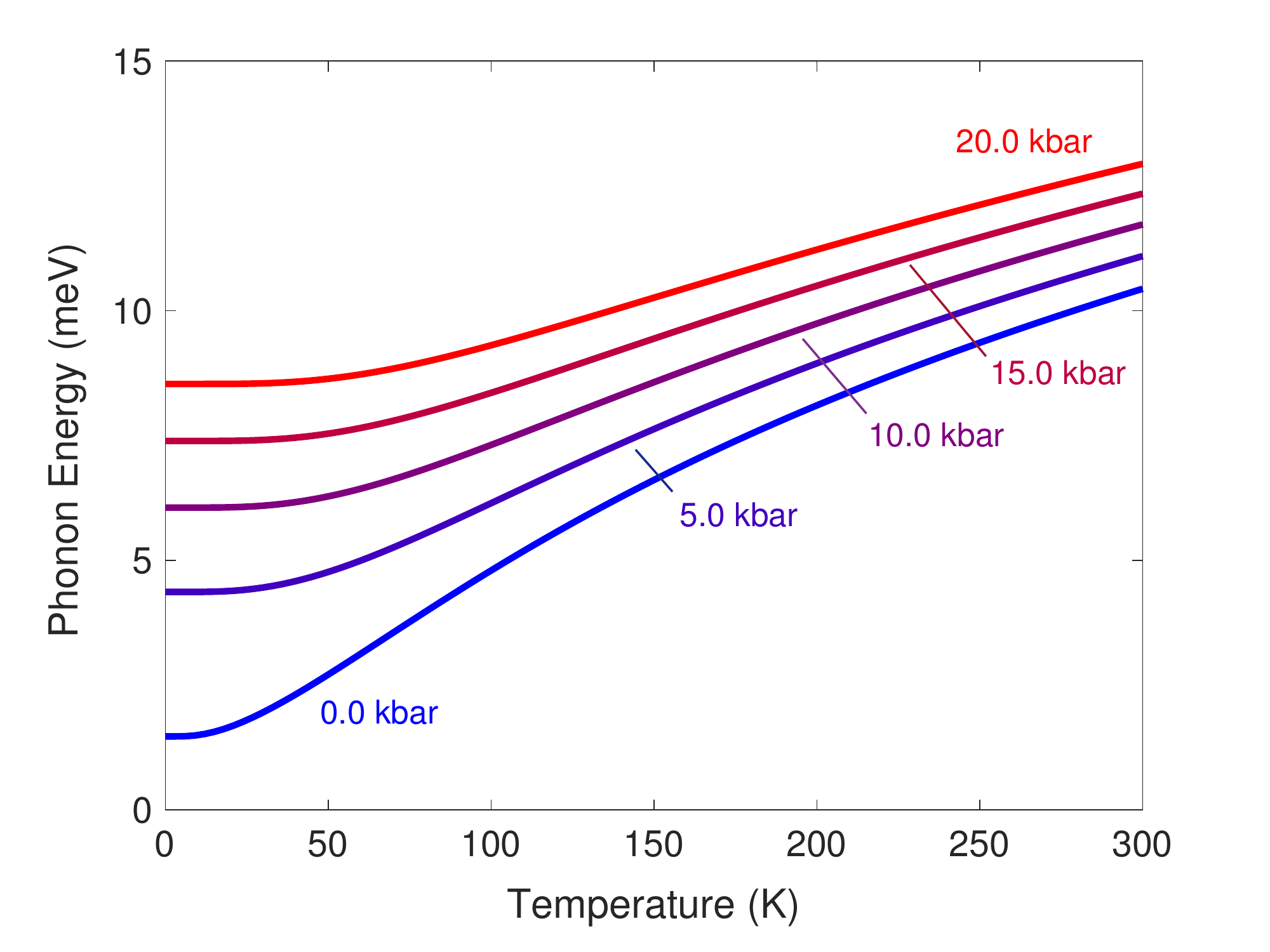}\caption{\label{fig:STO_Complex_PhononF_Plot}a) Zone-center soft mode phonon
energy of SrTiO$_3$ extracted from the dielectric constant via the Lydanne-Sachs-Teller
relation, plotted against temperature for applied pressures from ambient
(blue, lowest) to 20~kbar (red, topmost). The variation of the frequencies
of other phonon branches and the refractive index with temperature
and pressure were modeled as described in the text. Neutron scattering
data from Yamada et al. \citep{Yamada1969} are shown as open circles
and the corresponding data for KTaO$_3$ from Shirane et. al. \citep{Shirane1967}
as crosses. b) Calculated temperature dependence of the TO phonon energy for several pressures.}
\end{figure*}

\textbf{LST Analysis :} Such a drastic change in the dielectric properties must reflect a
change in the phonon spectrum of the system as the two are linked
by the LST - the phonon frequencies directly determine the dielectric
response. Eq. \ref{eq:LST_Reduced} was used to extract the temperature
and pressure dependence of the frequency $\omega_{TO}$ of the ferroelectric
soft mode $\omega_{TO}^{2}=\frac{\epsilon_{\infty}\omega_{LO}^{2}}{\epsilon_{r}}$
at the zone centre. $\epsilon_{\infty}$ and $\omega_{LO}$ only change
by \textasciitilde{}1\% over the pressure and temperature ranges studied,
as will be shown in the rest of this section, so a good approximation
of $\omega_{TO}$ can be found by simply setting these to be constant.
Including the known pressure and temperature dependencies of these
parameters will give a more accurate estimate of the phonon frequency
however, and this was done to find the data presented in Fig. \ref{fig:STO_Complex_PhononF_Plot}(a). 

The refractive index at high frequencies $\epsilon_{\infty}$ was
assumed to be temperature independent, in line with the conclusions
of Kamar{\'{a}}s et al. \citep{Kamaras1995}, but the pressure, $p$,
dependence was considered. The linear pressure dependence with a value
of $\left.\frac{\mathrm{d}\epsilon_{\infty}}{\mathrm{d}p}\right|_{300\,K}=-1.43\times10^{-4}\,\mathrm{kbar^{-1}}$
found by Giardini \citep{Giardini1957} for optical frequencies at
room temperature was used to model the changes in $\epsilon_{\infty}$.
An exact numerical value for $\epsilon_{\infty}$ is hard to define,
as $n$ exhibits frequency dependence out to the highest frequencies
measured \citep{Levin1955,Cardona1965}, but can be estimated as between
2.7 - 3.0 under ambient conditions. The numerical value for $\epsilon_{\infty}(p=0)$
was therefore set to 2.87 by fitting the extracted phonon frequencies
at ambient pressure to the experimental values found by Yamada et al.
\citep{Yamada1969} from inelastic neutron scattering.

The longitudinal optical mode frequency $\omega_{LO}$ has been found
to be $150\,\mathrm{cm}^{-1}=18.6\,\mathrm{meV}$ under ambient conditions
by Ishidate et al. \citep{Ishidate1992} and Lebedev \citep{Lebedev2009}.
Ishidate measured room temperature phonon frequencies in SrTiO$_3$ up
to a pressure range an order of magnitude higher than that reported
in this work, and found a linear pressure dependence throughout, with
$\left.\frac{\mathrm{d}\omega_{LO}}{\mathrm{d}p}\right|_{300\,K}=1.8\,\mathrm{cm^{-1}}\mathrm{GPa}^{-1}=0.0223\,\mathrm{meV\,kbar}^{-1}$.
The temperature variation of $\omega_{LO}$ at ambient pressure was
reported by Servoin et al. \citep{Servoin1980}, and this was fitted
with a linear relationship to extract a gradient $\left.\frac{\mathrm{d}\omega_{LO}}{\mathrm{d}T}\right|_{p=0}=0.0066\,\mathrm{cm^{-1}}\mathrm{K}^{-1}=0.00082\,\mathrm{meV\,K}^{-1}$.
Both pressure and temperature effects are therefore on order a percent
over the ranges studied, just as with the refractive index changes,
so all three must be considered for a rigorous analysis.

Holding $\omega_{LO}$ and $\epsilon_{\infty}$ constant as in Yamada
et al. and fitting the resulting constant of proportionality to match
the Yamada data gives $\omega_{TO}^{2}=\frac{39204}{\epsilon_{r}}$,
with frequencies $\omega$ in meV and $\epsilon_{r}=\epsilon/\epsilon_{0}$
dimensionless. Yamada had a similar conversion factor of 37790, suggesting
comparable sample quality - the dielectric constant is very sensitive
to sample purity and surface finish, easily explaining the slight
difference. Including the variation in $\omega_{LO}$ and $\epsilon_{\infty}$
leads to: 
\[
\omega_{TO}^{2}=\frac{(2\pi)^{2}}{\epsilon_{r}}\left[2.87-\frac{p}{6993}\right]\left[18.6+\frac{p}{44.8}+\frac{(T-300)}{1220}\right]
\](pressures in kbar, temperature in K). 
\\

\textbf{Phonon Frequency Results :} This last equation was used to extract the
values for $\omega_{TO}$ shown in Fig. \ref{fig:STO_Complex_PhononF_Plot}(a).
As pressure is increased the phonon frequency markedly flattens off
and returns towards the flat temperature dependence of an archetypal
optical mode. The low-temperature gap, $\Delta$, is doubled from
2~meV by even 2.4~kbar, the smallest pressure applicable with this
setup. The small phonon energy or gap is directly linked to the ferroelectric
quantum critical fluctuations observed in this material, so the gap
widening as seen and becoming energetically inaccessible as the system
is tuned away from criticality is an expected result. 
The prediction that pressure will tune the system towards the behavior seen in KTaO$_3$
is supported by the data as well - the soft mode energy for KTaO$_3$
overlayed on Fig. \ref{fig:STO_Complex_PhononF_Plot}(a) lie by inspection
at where SrTiO$_3$ values at around 2~kbar would be found.

\begin{figure*}
\begin{raggedright}
a)\hspace{1.0\columnwidth}b)
\par\end{raggedright}
\centering{}\includegraphics[width=0.4\paperwidth]{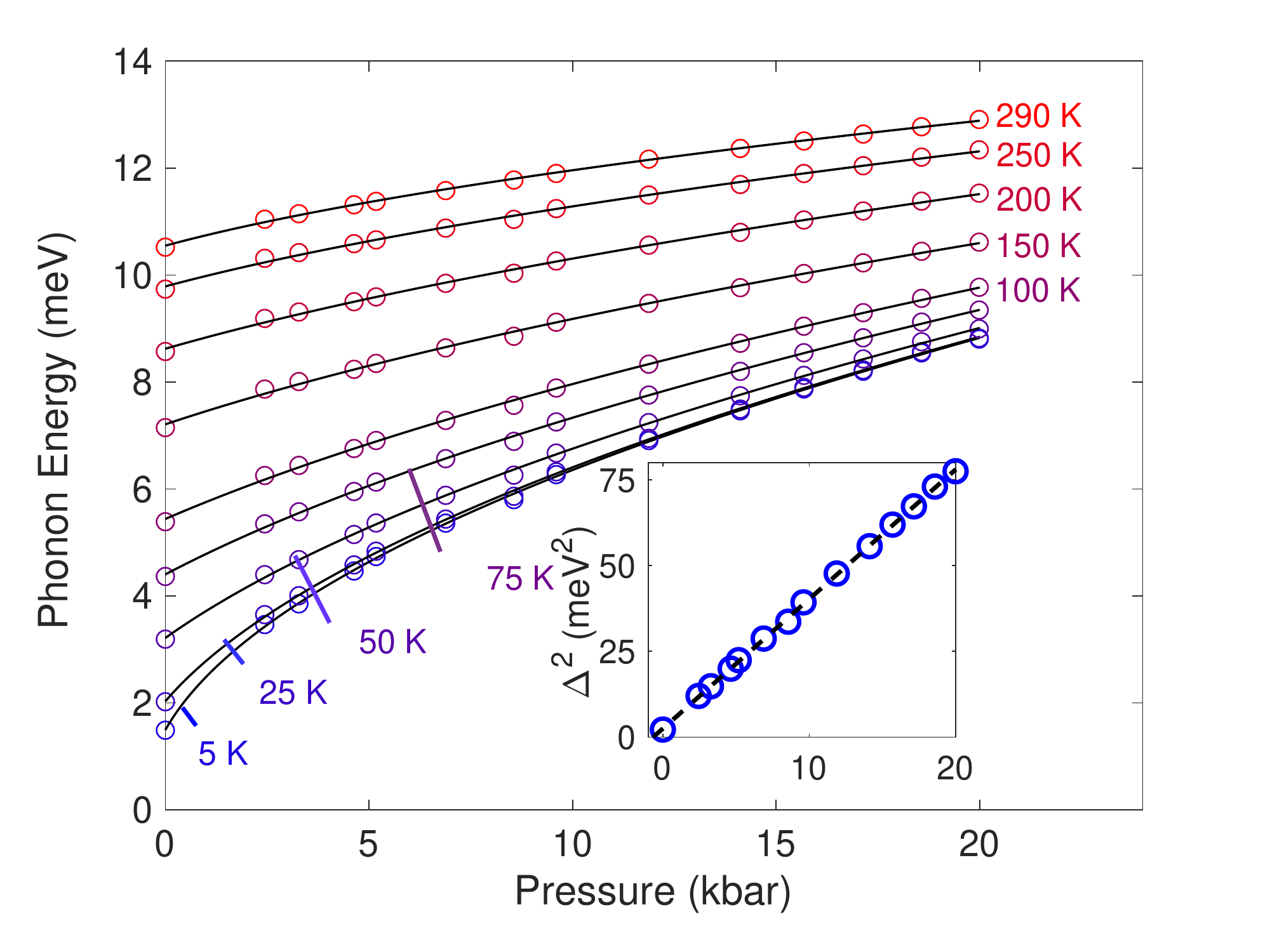}\hfill{}\includegraphics[width=0.4\paperwidth]{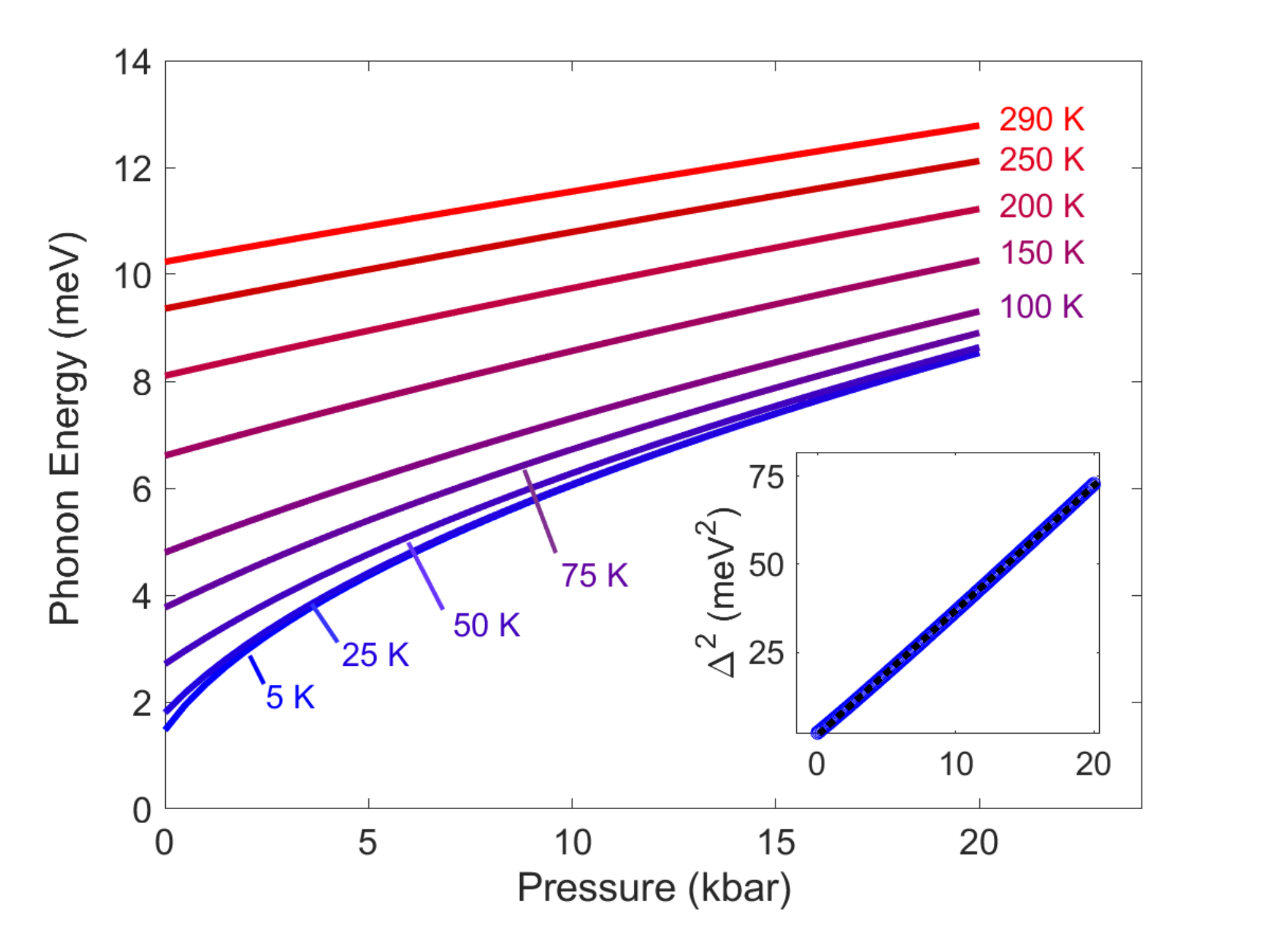}\caption{\label{fig:STO_Complex_PhononF_FixedTemperatures}a) Zone-center soft
mode phonon energy of SrTiO$_3$, plotted against pressure at fixed temperatures.
The solid lines show fits to $\omega_{0}(p-p_{c})^{0.5}$ with $\omega_{o}$
and $p_{c}$ free fitting parameters to guide the eye. The inset gives
the square of the zero-temperature extrapolated phonon gap,
$\Delta$, against pressure, with a linear fit shown as a solid line. b) Calculated pressure dependence of the TO phonon energy for several temperatures. Inset. Calculated pressure dependence of the TO phonon gap.}
\end{figure*}

Fig. \ref{fig:STO_Complex_PhononF_FixedTemperatures}(a) shows the pressure dependence of $\omega_{TO}$ at fixed temperatures, taken directly from the data of Fig. \ref{fig:STO_Complex_PhononF_Plot}(a). The data at 1.5~K, 2~K, 5~K and those extrapolated to 0~K all overlap as the dielectric response flattens off as temperature approaches zero, so any data in this range can be taken as the zero-temperature value. The inset of Fig. \ref{fig:STO_Complex_PhononF_FixedTemperatures}(a) shows the square of the phonon energy at 2~K, equivalent to the zero-temperature gap, $\Delta$, plotted against pressure. This shows very good agreement to a linear fit, giving the behavior expected from Landau theory $\omega=\omega_{0}(p-p_{c})^{0.5}$ with the critical pressure $p_{c}$ at which the ferroelectric quantum critical point occurs equal to -0.7(1)~kbar. This is in agreement with the analysis in our earlier paper \citep{Coak2018} and with observations that strain or oxygen-18 substitution (both equivalent to `negative pressure' in SrTiO$_3$) drive the system to a ferroelectric state \citep{Uwe1976,Itoh2000a,Wang2000}.

\section{Theory}

\subsection{Model}

We consider a standard 
model Hamiltonian for displacive ferroelectrics
with normal mode coordinates that describe local displacements 
$ \phi_{i \alpha}$, with $\alpha=1,2,...,n=d$ in the unit cell $i=1,2,...N$  that are 
associated with the soft TO mode, the condensation of which is driven by 
the dipolar force and leads to the ferroelectric transition. We also consider 
the coupling to volume strain $\eta_i$.
The  Hamiltonian is as follows,~\citep{Arce-Gamboa2018a}
\begin{align}
	\label{eq:Hamiltonian}
	H = H_\phi + H_\eta + H_{\phi \eta},
\end{align}
where $H_\phi$ is the Hamiltonian of the polar degrees of freedom alone,
\begin{align*}
H_\phi &=    \frac{1}{2} \sum_{i, \alpha  }    \Pi_{ i \alpha}^2 + \frac{\omega_0^2}{2} \sum_{ i, \alpha }   \phi_{i \alpha }^2 
+ \sum_{i, \alpha \beta} \left[ u + v \delta_{\alpha \beta} \right] \phi_{ i \alpha }^2  \phi_{ i \beta }^2 \nonumber \\
&
+\frac{1}{2} \sum_{ij, \alpha \beta} F_{\alpha \beta} \left( i-j \right)  \phi_{ i \alpha }   \phi_{ j \beta },  
\end{align*}
$H_\eta$ is the elastic energy,
\begin{align*}
H_\eta &=  \frac{K}{2} \sum_i  \eta_i^2  + P \sum_i \eta_i,
\end{align*}
and $H_{\phi\eta}$ is a linear-quadratic coupling between $\phi_{i \alpha}$ and $\eta_i$,
\begin{align*}
H_{\phi\eta}  &=  -  \lambda  \sum_{ i, \alpha } \eta_{i}  \phi_{ i \alpha }^2. 
\end{align*}
%
Here, $ \Pi_{i \alpha} $ is  
the conjugate momentum of $ \phi_{i \alpha}  $;  and 
$ F_{\alpha\beta}  \left( i-j \right)  $ is the dipolar interaction with 
Fourier transform
$
F_{ \alpha \beta} \left( {\bm q} \right) 
=  q^2   \delta_{ \alpha \beta } + g_0  q_{ \alpha } q_{ \beta } / q^2 ,  
$
where  $g_0$ 
is a constant that depends on the lattice structure.
$\omega_0$ is the frequency of the purely harmonic model; $u$ and $v$ are 
coefficients of
the isotropic and anisotropic cubic anharmonicities, respectively;
$\lambda$ is the coupling constant between the polarization and volume strain; and 
$K$ and $P$ are, respectively, the  bulk modulus and applied hydrostatic stress~(e.g. biaxial stress for $d=2$ and pressure for $d=3$). 

Though the low temperature paraelectric lattice structure of SrTiO$_3$ is tetragonal, we will follow previous work \citep{Rowley2014} and assume cubic symmetry for simplicity. The TO excitations of  (\ref{eq:Hamiltonian}) are thus doubly degenerate,
\begin{align}
\label{eq:TOLOC}
\Omega_{ {\bm q}}^2 =  r(T,P) + q^2,
\end{align}
where $r(T,P) \equiv \Delta^2(T,P)$ is the square of the phonon gap at the zone-center.
In addition to (\ref{eq:TOLOC}), there is  a singlet LO mode which 
 we will neglect as it is gapped out by the large depolarizing field of order $\mathcal{O}\left( g \right)$.  Within the SCPA,~\citep{Arce-Gamboa2018a, Arce-Gamboa2017a} $r(T,P)$  is given as follows,
\begin{align}
\label{eq:TOLOC0}
r(T,P)  &=   \omega_0^2 + \left[   4\left( d+2 \right) u + 12 v  \right]  \bra \left| {\bm \phi} \right|^2 \ket - 2 \lambda \bra \eta \ket,
\end{align}
where   $ \bra \left| {\bm \phi} \right|^2 \ket$ are the  fluctuations of polarization,
\begin{align}
\label{eq:Psi0}
\bra \left| {\bm \phi} \right|^2 \ket 
=   \frac{1}{N} \sum_{\bm q}   \frac{ 1 }{\Omega_{\bm q} }  \left[ n\left(   \Omega_{\bm q} / k_B T \right) + \frac{1}{2}   \right], 
\end{align}
with $n(\xi) = \left( e^{\xi} - 1 \right)^{-1}$ a Bose factor and 
 $\bra \eta \ket$ is the average volume strain,
\begin{align}
\label{eq:eta_a}
\bra \eta \ket = \frac{\lambda}{K}  \bra \left| {\bm \phi} \right|^2 \ket  - \frac{P}{K}.
\end{align}

By substituting Eqs.~(\ref{eq:Psi0}) and~(\ref{eq:eta_a}) into Eq.~(\ref{eq:TOLOC0}), we 
separate the contributions from thermal and quantum fluctuations as follows~\cite{Schneider1976},
\begin{align}
\label{eq:rTP}
r(T,P) &= \frac{2 \lambda}{K} \left( P - P_c \right) \\ 
&+ \Gamma \left\{ \left(T/T_D \right)^{d-1} \Phi\left[ r\left( T, P\right) /  \left( k_B T \right)^2 \right] \right. \nonumber \\
&~~~~~~~~~~~~~~~~~~~~~~~ \left. + Z_0 \left[ r\left(0, P\right) \right] - Z_0\left[ 0 \right]  \right\}, \nonumber
\end{align}
where $P_c$ is the critical pressure at $T=0\,$K,
\begin{align}
	\label{eq:Pc}
\frac{2 \lambda}{K} P_c = - \left[ r(0,0) + \Gamma  \left\{ Z_0 \left[0 \right] - Z_0\left[  r\left(0, 0\right) \right]  \right\} \right],
\end{align}
and, 
\begin{subequations}
	\label{eq:Z}
	\begin{align}
		\Phi\left[ R \right] &=  \frac{\left( k_B T_D\right)^{d-1}}{\left(2 \pi \right)^d} \nonumber \\ 
		& ~~~\times \int_0^{T_D/T} dx x^{d-1} \frac{1}{\sqrt{R + x^2}} n\left( \sqrt{R + x^2} \right),\\
Z_0(r) &=  \frac{\left(k_B T_D\right)^{d-1}}{2\left(2 \pi \right)^d} \int_0^1 dx x^{d-1}  \frac{1}{\sqrt{ r/ \left( k_B T_D \right)^2 +  x^2 }}, 
\end{align}
\end{subequations}
where $ r(0,0) = \omega_0^2 + \Gamma Z_0 \left[ r\left(0, 0\right)  \right]$, $ \Gamma =   4\left( d+2 \right) u  + 12 v   - 2 \lambda^2/K   $, and  $ T_D = \Lambda / k_B $ is the Debye temperature.
Eqs.~(\ref{eq:rTP}), (\ref{eq:Pc}), and (\ref{eq:Z})  self-consistently determine the temperature and pressure dependence of $r\left(T,P\right)$.

\subsection{Results}

The non-adjustable model parameters are $\omega_0, \Gamma, \lambda, K$, and $\Lambda$
which we fit experimental data reported for SrTiO$_3$ as follows:
$\omega_0$ and $ \Gamma$ are determined from the temperature dependence of the transverse optic phonon gap measured by neutron scattering,~\cite{Yamada1969};
$ \lambda / K$ is obtained from the slope of $\Delta^2$ versus $P$ measured here (see inset of Fig.~\ref{fig:STO_Complex_PhononF_FixedTemperatures}(b));
and $\Lambda$ from specific heat data~($T_D = \Lambda / k_B = 400\,$K) ~\cite{Burns1980}. 
The resulting parameters are $ \omega_0 = 10.2 i\,$meV, $\Gamma = 90.0 \,$meV$^3$, $\lambda / K = 0.054\,$meV$^2$,
and $\Lambda = 34.5\, $meV.  The calculated temperature and pressure dependence of $r(T,P)$ is in 
good qualitative and quantitative agreement with experiments,  as shown in Figs.~\ref{fig:STO_Complex_PhononF_Plot} and \ref{fig:STO_Complex_PhononF_FixedTemperatures}. 

To further explore the behavior of the excitations, 
we study the pressure dependence of 
the fluctuations near the QCP. At $T=0\,$K, Eq.~(\ref{eq:rTP}) gives,
\begin{align}
\label{eq:rTP@0K}
r(0,P) &= \frac{2 \lambda}{K} \left( P - P_c \right) + \Gamma \left\{ Z_0 \left[ r\left(0, P\right) \right] - Z_0\left[ 0 \right]  \right\}.
\end{align}
For $r \to 0$,
\begin{align}
\label{eq:Z0@0K}
Z_0\left[r\right] - Z_0\left[0 \right]   \simeq \begin{cases} 
- b \ln r + c r, & d=1, \\ 
 - b r^{1/2} + c r , & d=2, \\
 - b r + c r \ln r , & d=3, \\
 - b r +  c r^{3/2}, & d = 4, \\
\end{cases} 
\end{align}
with $b,c$ positive constants.
Thus the lower (upper) critical dimension is $d_{lc}=1~(d_{uc}=4)$. 
While there are strong deviations from mean field behavior for $d=2$,
the corrections are only logarithmic for $d=3$. 
Equation.~(\ref{eq:rTP@0K}) then gives, 
\begin{align}
	\label{eq:r@0K}
	r(0,P) \propto \begin{cases} \left( P - P_c \right)^2, & d =2,  \\ \left( P - P_c \right), & d \geq 3, \end{cases}
\end{align}
with weak logarithmic corrections for $d=3$. Such corrections
should be observable only when extremely close to the QCP ($\left| \left(P - P_c \right) / P_c \right| \ll 1$) and a renormalization group analysis would be required for a quantitative description. 
For the present experiments, $\left| \left(P - P_c \right) / P_c \right| \gtrsim 1$, and thus are not observable.

Quantum critical excitations in uniaxial and $XY$ ferroelectrics are generally expected to differ from those of cubic systems. We briefly consider such cases in Appendix A.

\section{Conclusions}

 The Lydanne-Sachs-Teller relation was used to extract the pressure and temperature dependence of the transverse optical soft mode frequency in SrTiO$_3$ from precision measurements of the dielectric constant under pressure. The mode was seen to stiffen as pressure was applied, moving towards a more typical flat optical mode. The low-temperature phonon gap $\Delta$ widened with increasing pressure, showing excellent agreement to the $\Delta\propto(p-p_{c})^{0.5}$ predicted by Landau theory and with an extrapolated critical pressure $p_{c}$ of -0.7(1)~kbar.

By treating the pressure within a self-consistent model for the quantum ferroelectric excitations, we have shown that the pressure dependence is robustly mean-field like. There are logarithmic corrections very close to absolute zero but these extend only to very low pressures (much less than 1 kbar). At the same time we verify the applicability of the LST over the whole of the phase space considered.

SrTiO$_3$ forms an ideal model system for the study of the physics of soft phonon modes and of quantum criticality, and pressure allows direct and clean tuning of the phonons - the driving mechanism of the whole system. Our results provide the first measurement of the stiffening of a soft phonon mode as a system is tuned away from criticality in a clean and direct manner by the application of hydrostatic pressure - a potentially universal phenomenon across a variety of phase transitions and systems in condensed matter physics.

\section*{Acknowledgements} 
\begin{acknowledgments}
The authors would like to thank G.G. Lonzarich, P.B. Littlewood, F.M.
Grosche, D.E. Khmelnitskii, S.E. Dutton, N.D. Mathur, J. van Wezel,
G. Tsironis, C. Panagopoulos, L.J. Spalek, H. Hamidov, P.A.C. Brown, D.
Jarvis and J. Zaanen for their help and discussions. We would also
like to acknowledge support from Jesus, Churchill and Trinity Colleges of the
University of Cambridge, the Engineering and Physical Sciences Research
Council, IHT  KAZATOMPROM and the CHT programme. The work was carried
out with financial support from the Ministry of Education and Science
of the Russian Federation in the framework of Increase Competitiveness
Program of NUST MISiS (\textnumero{} K2-2017-024), implemented
by a governmental decree dated 16th of March 2013, N 211. 
Work at the University of Costa Rica is supported by the
Vice-rectory for Research under the project no. 816-B7-601 and 
the Office of International Affairs.
\end{acknowledgments}

\appendix

\section{Uniaxial and $XY$ ferroelectrics}

For uniaxial ferroelectrics, there is an  preferred polarization direction which we choose to be the $x_3$-axis. The Hamiltonian is similar to that of Eq.~(\ref{eq:Hamiltonian}) except that $\alpha=\beta=3, v=0$, and $F({\bm q})=q^2 + g \left( q_3/ q \right)^2$. 
The phonon excitations are $\Omega_{\bm q} = \sqrt{ r+ q^2 + g  \left( q_3/ q \right)^2}$,~\cite{Roussev2003} which give
\begin{align}
	Z_0\left[r\right] - Z_0\left[0 \right]   = 
	\begin{cases}   b r + c r \ln r  + \mathcal{O}\left( r^2 \right), & d=2, \\
						   b r    + \mathcal{O}\left( r^{3/2} \right), & d=3.
	\end{cases}
\end{align}
The upper critical dimension is thus $d_{uc}=3$ and there are weak logarithmic corrections
in $d=2$.

For $XY$ ferroelectrics, there is an easy plane of polarization and the Hamiltonian is again similar to that of Eq.~(\ref{eq:Hamiltonian}) except that $\alpha=\beta={1,2} $. 
The excitations are 
$\Omega_{\bm q} = \sqrt{ r +  q^2 }$ and $\Omega_{\bm q} = \sqrt{ r+  q^2 + g  \left(q_1^2 + q_2^2\right)/ q^2}$. The former isotropic dispersion dominates the critical behavior and the corrections to mean-field behavior are thus the same as those given in Eq.~(\ref{eq:Z0@0K}).


\begin{thebibliography}{35}

\bibitem{Barrett1952} 
J. H. Barrett, Dielectric constant in perovskite type crystals, {Physical Review} \textbf{86}, 118-120 (1952)   

\bibitem{Cowley1964} 
R. A. Cowley, Lattice dynamics and phase transitions of strontium titanate, {Physical Review} \textbf{134}, A981-A997 (1964) 

\bibitem{Samara1966} 
G. A. Samara, Pressure and temperature dependences of the dielectric properties of the perovskites {BaTiO}$_3$ and {SrTiO}$_3$, {Physical Review} \textbf{151}, 378-386 (1966) 

\bibitem{Mueller1979} 
K. A. M\"{u}ller and H. Burkard, {SrTiO}$_3$ : An intrinsic quantum paraelectric below 4~{K}, {Phys. Rev. B} \textbf{19}, 3593-3602 (1979)  

\bibitem{Muller1991} 
K. A. M\"{u}ller, W. Berlinger and E. Tosatti, Indication for a novel phase in the quantum paraelectric regime of {SrTiO}$_3$, {Z. Physik B - Condensed Matter} \textbf{84}, 277-283 (1991) 

\bibitem{Fujishita2016} 
H. Fujishita, S. Kitazawa, M. Saito, R. Ishisaka, H. Okamoto and T. Yamaguchi, Quantum Paraelectric States in {SrTiO}$_3$ and {KTaO}$_3$: {Barrett Model, Vendik Model, and Quantum Criticality}, {J. Phys. Soc. Jpn.} \textbf{85}, 074703 (2016)  

\bibitem{Atkinson2017} 
W. A. Atkinson, P. Lafleur and A. Raslan, Influence of the ferroelectric quantum critical point on {SrTiO}$_3$ interfaces, {Phys. Rev. B} \textbf{95}, 054107 (2017) 

\bibitem{Coak2018} 
M. J. Coak, C. R. S. Haines, C. Liu, S. E. Rowley, G. G. Lonzarich and S. S. Saxena, Emergence of a quantum coherent state at the border of ferroelectricity, {ArXiv preprint} 1808.02428, (2018) 

\bibitem{Shirane1969} 
G. Shirane and Y. Yamada, Lattice-dynamical study of the 110 {K} phase transition in {SrTiO}$_3$, {Physical Review} \textbf{177}, 858-863 (1969)  

\bibitem{Fleury1968} 
P. A. Fleury and J. M. Worlock, Electric-field-induced {R}aman scattering in {SrTiO}$_3$ and {KTaO}$_3$, {Physical Review} \textbf{174}, 613-623 (1968)  

\bibitem{Yamada1969} 
Y. Yamada and G. Shirane, Neutron scattering and nature of the soft optical phonon in {SrTiO}$_3$, {J. Phys. Soc. Jpn.} \textbf{26}, 396-403 (1969)  

\bibitem{Uwe1976} 
H. Uwe and T. Sakudo, Stress-induced ferroelectricity and soft phonon modes in {SrTiO}$_3$, {Phys. Rev. B} \textbf{13}, 271-286 (1976)  

\bibitem{Itoh2000a} 
M. Itoh and R. Wang, Quantum ferroelectricity in {SrTiO}$_3$ induced by oxygen isotope exchange, {Applied Physics Letters} \textbf{76}, 221-223 (2000)

\bibitem{Wang2000} 
R. Wang, N. Sakamoto and M. Itoh, Effects of pressure on the dielectric properties of {SrTi}$^{18}${O}$_3$ and {SrTi}$^{16}${O}$_3$ single crystals, {Phys. Rev. B} \textbf{62},  R3577 (2000). 

\bibitem{Coak2018b} 
M. J. Coak, C. R. S. Haines, C. Liu, D. M. Jarvis, P. B. Littlewood and S. S. Saxena, Dielectric response of quantum critical ferroelectric as a function of pressure, {Sci. Rep.} \textbf{8}, 14936 (2018)  

\bibitem{Lonzarich_MagneticElectron} 
G. G. Lonzarich, The Magnetic Electron, {Cambridge University Press} (1997)   

\bibitem{LinesAndGlass} 
M. A. Lines and A. M. Glass, Principles and applications of ferroelectrics and related materials, {Clarendon Press} (1970) 

\bibitem{Axe1970} 
J. Axe, J. Harada and G. Shirane, Anomalous acoustic dispersion in centrosymmetric crystals with soft optic phonons, {Phys. Rev. B} \textbf{1}, 1227-1234 (1970)  

\bibitem{Rowley2014} 
S. E. Rowley, L. J. Spalek, R. P. Smith, M. P. M. Dean, M. Itoh, J. F. Scott, G. G. Lonzarich and S. S. Saxena, Ferroelectric quantum criticality, {Nat. Phys.} \textbf{10}, 367-372 (2014)  

\bibitem{Arce-Gamboa2017a} 
J. R. Arce-Gamboa and G. G. Guzm{\'{a}}n-Verri, Random electric field instabilities of relaxor ferroelectrics, {npj Quantum Materials} \textbf{2}, 28 (2017)

\bibitem{Arce-Gamboa2018a} 
J. R. Arce-Gamboa and G. G. Guzm{\'{a}}n-Verri, Quantum ferroelectric instabilities in superconducting SrTiO$_3$, {Phys. Rev. Mat.} \textbf{2}, 10484 (2018)

\bibitem{Loehneysen2007} 
H. v. L{\"o}hneysen, A. Rosch, M. Vojta and P. W{\"o}lfle, Fermi-liquid instabilities at magnetic quantum phase transitions, {Reviews of Modern Physics} \textbf{79}, 1015-1075 (2007)  

\bibitem{Lyddane1941} 
R. H. Lyddane, R. G. Sachs and E. Teller, On the polar vibrations of alkali halides, {Physical Review} \textbf{59}, 673-676 (1941)

\bibitem{Barker1966} 
A. S. Barker, Temperature dependence of the transverse and longitudinal optic mode frequencies and charges in SrTiO$_3$ and BaTiO$_3$, {Physical Review} \textbf{145}, 391-399 (1966)  

\bibitem{Guennou2010} 
M. Guennou, P. Bouvier, J. Kreisel and D. Machon, Pressure-temperature phase diagram of {SrTiO}$_3$ up to 53~{GPa}, {Phys. Rev. B} \textbf{81}, 054115 (2010)

\bibitem{Ishidate1992} 
T. Ishidate and T. Isonuma, Phase transition of {SrTiO}$_3$ under high pressure, {Ferroelectrics} \textbf{137}, 45-52 (1992) 

\bibitem{Beattie1971} 
A. G. Beattie, Pressure dependence of the elastic constants of {SrTiO}$_3$, {Journal of Applied Physics} \textbf{42}, 2376 (1971)

\bibitem{Shirane1967} 
G. Shirane, R. Nathans and V. J. Minkiewicz, Temperature dependence of the soft ferroelectric mode in {KTaO}$_3$, {Physical Review} \textbf{157}, 396-399 (1967)  

\bibitem{Kamaras1995} 
K. Kamar{\'{a}}s,  K. -L. Barth, F. Keilmann, R. Henn, M. Reedyk, C. Thomsen, M. Cardona, J. Kircher, P. Richards and J. -L. Stehl{\'{e}}, The low-temperature infrared optical functions of {SrTiO}$_3$ determined by reflectance spectroscopy and spectroscopic ellipsometry, {Journal of Applied Physics} \textbf{78}, 1235-1240 (1995)  

\bibitem{Giardini1957} 
A. A. Giardini, Stress-optical study of strontium titanate, {Journal of the Optical Society of America} \textbf{47}, 726 (1957) 

\bibitem{Levin1955} 
S. B. Levin, N. J. Field, F. M. Plock and L. Merker, Some optical properties of strontium titanate crystal, {J. Opt. Soc. Am.} \textbf{45}, 737-739 (1955)

\bibitem{Cardona1965} 
M. Cardona, Optical properties and band structure of {SrTiO}$_3$ and {BaTiO}$_3$, {Physical Review} \textbf{140}, A651-A655 (1965) 

\bibitem{Lebedev2009} 
A. I. Lebedev, Ab initio calculations of phonon spectra in {ATiO}$_3$ perovskite crystals {(A = Ca, Sr, Ba, Ra, Cd, Zn, Mg, Ge, Sn, Pb)}, {Physics of the Solid State} \textbf{51}, 362-372 (2009)

\bibitem{Servoin1980} 
J. L. Servoin, Y. Luspin and F. Gervais, Infrared dispersion in SrTiO$_3$ at high temperature, {Phys. Rev. B} \textbf{22}, 5501-5506 (1980)

\bibitem{Schneider1976} 
T. Schneider, H. Beck and E. Stoll, Quantum effects in an n-component vector model for structural phase transitions, {Phys. Rev. B} \textbf{13}, 1123-1130 (1976) 

\bibitem{Burns1980} 
G. Burns, Comment on the low temperature specific heat of ferroelectrics, antiferroelectrics, and related materials, {Solid State Comm.} \textbf{35}, 811-814 (1980)

\bibitem{Roussev2003} 
R. Roussev and A. J. Millis, Theory of the quantum paraelectric-ferroelectric transition, {Phys. Rev. B} \textbf{67}, 014105 (2003) 

\end{thebibliography}
\end{document}